\documentclass[english,aps,pre,twocolumn,showpacs,superscriptaddress,floatfix]{revtex4-1}
\usepackage[T1]{fontenc}
\usepackage[latin1]{inputenc}
\usepackage{amsmath}
\usepackage{graphicx}
\usepackage{subfig}
\usepackage{amssymb}
\usepackage{dcolumn} 
\usepackage{bm} 
\usepackage{color}
\usepackage{cleveref}

\makeatletter

\usepackage{amsfonts}
\usepackage{epsfig}
\usepackage{dsfont}

\newcommand{\be}{\begin{equation}}
\newcommand{\bea}{\begin{eqnarray}}
\newcommand{\eea}{\end{eqnarray}}
\newcommand{\ee}{\end{equation}}

\usepackage{babel}
\makeatother

\begin{document}

\title{Hysteresis of economic networks in an XY model}

\author{Ali \surname {Hosseiny}}\email{Al_hosseiny@sbu.ac.ir}
\affiliation{Department of Physics, Shahid Beheshti University, G.C., Evin, Tehran 19839, Iran} 
\author{Mohammadreza \surname {Absalan}}
\affiliation{Department of Physics, Sharif University of Technology \\
P.O. Box 11165-9161, Tehran, Iran. }
\author{Mohammad \surname {Sherafati}}
\affiliation{Department of Physics, Sharif University of Technology \\
P.O. Box 11165-9161, Tehran, Iran. }
\author{Mauro \surname {Gallegati}}
\affiliation{Department of Economics, Universit\`a Politecnica delle Marche, Italy}

\date{\today}



\begin{abstract}



Many-body systems can have multiple equilibria. Though the energy of equilibria might be the same, still systems may resist to switch from an unfavored equilibrium to a favored one.
In this paper we investigate occurrence of such phenomenon in economic networks. In times of crisis when governments intend to stimulate economy, a relevant question is on the proper size of stimulus bill. To address the answer, we emphasize the role of hysteresis in economic networks. In times of crises, firms and corporations cut their productions; now since their level of activity is correlated, metastable features in the network become prominent. This means that economic networks resist against the recovery actions. To measure the size of resistance in the network against recovery, we deploy the XY model. Though theoretically the XY model has no hysteresis, when it comes to the kinetic behavior in the deterministic regimes, we observe a dynamic hysteresis. 
We find that to overcome the hysteresis of the network, a minimum size of stimulation is needed for success. 
Our simulations show that as long as the networks are Watts-Strogatz, such minimum is independent of the characteristics of the networks.

\end{abstract}


\maketitle
\thispagestyle{empty}
\section*{Introduction}

The role of network and its structure has proved to be crucial in addressing a wide range of phenomena in complex systems. In economics it has been shown that the structure of the  network can influence fragility of the market \cite{Schweitzer}-\cite{contreras}. As well it has been proved that despite the classical view where the fluctuations in a regular network may cancel out, in scale free networks such fluctuations may contribute to a turn over of the market \cite{Acemoglu}.

From a physical perspective and in a simplified world, economy can be viewed as a network of agents which interact with each other. When agents interact in a system and try to maximize a function such as utility, then we expect to have a wide range of local equilibria. Existence of a spectrum of local equilibria then makes it hard to derive the system to a favored equilibrium and results in a hysteresis in the network. In this paper we notify such hysteresis in economic network in the context of the XY model. 

A hot issue in economics which was rerose after the crash of the Lehman Brothers was the role of government in the time of crisis and stimulating policies. Obama's stimulus polices along with the Federal Reserve expansionary programs were successful to help the economy of the United States for recovery. In the debates concerning the stimulus programs, some economists claimed that only big stimulations could help economy for a fast recovery, see for example \cite{{krugmanend},{stiglitzfreefall}}. In other words, there was an inspiration that market may resist for recovery.

Existence of such resistance has been studied in \cite{hosseinyising} in an Ising based model of the network of firms. Actually firms and corporations are customers of each other's products. They buy and sell their products to each other as intermediate goods. Now we can consider a network in which nodes are firms or corporations. Two nodes are connected if they have economic interactions. In other words two firms are connected if they trade. 

When two firms are connected, there should be some positive correlations between their level of activities. If a firm increases/decreases it production, it buys or sells more/less from its neighbors. As a result it forces its neighbors to work with higher/lower level of activity. This positive correlation provides a hysteresis in the network.

In the United States after the crisis of 2008 different states faced the crisis with different levels. Consequently, the rate of recovery was different after the crisis. In April 2012 while unemployment was as low as five percent in Iowa, it was close to eleven percent in California. When you provide a service in California you cannot hire new employees if economy is good in Iowa or other states. You should look at the activities of your neighbors in the network.

 As a result of local interaction and positive correlations between neighbors, and similar to many ferromagnet models, we might have global order in the system and consequently resistance for change in the global states. Ising model as a model to explain the behavior of manager, firms, and corporation has been presented before, see \cite{William}, \cite{Durlauf}. It as well has been utilized to model behaviors in the financial markets \cite{Zhou}. Actually, in the simplest model we can suppose that firms have the choice to work with either their maximum capacity or their minimum capacity. So, they have binary choices. We then can think what happens when we want to stimulate firms which work with minimum capacity. We should overcome the effect of neighbors and bypass the wall between two vacuums of the system. Studying such problem we find that there is a minimum cost to change states of an Ising model. In other words stimulations with a cost below that threshold fails to recover economy. It seems that it is mainly because in the spectrum of energies of the configuration of states, there is a hump between two vacuums of the Ising model.

Simulating the network of firms with an Ising model sheds light on the metastable features of networks and hysteresis against recovery. It has the benefit of being simple to model, analyze, and simulate. It has however its own restrictions. The major problem is that firms have much more than a binary choice. Basically they can work with any level of activities within a range. Now, one relevant question is what happens if we consider firms with a continuos choice of activity. Does there still exist a minimum bound for successful stimulations? To answer this question we simulate network of firms with an XY model. Despite the Ising model, in the XY model, each agent can have a continuos choice. As a result in the configuration space, vacuums can change without a hump in energy. Now, it is interesting to check if going from an agent base model with binary choices to another one with continuous choice still for successful stimulation we need a minimum of cost.



\section*{Hysteresis in an Ising Model}

Consider two firms or corporations which trade with each other. We connect them in network of firms via an edge. In each firm, managers can rise or reduce working hours. Further, they can employ new or fire old employees. In other words they have choice to decrease or increase the level of their productions or services. This choice however is limited. There is maximum capacity where production above it is impossible, and a minimum capacity where production below it results in loss. In the simplest case we suppose that firms have a binary choice of working with either maximum or minimum level of activity, see \cite{William}. 
%
%
%
%
 
\begin{figure}[]
  \includegraphics[width=.8\columnwidth]{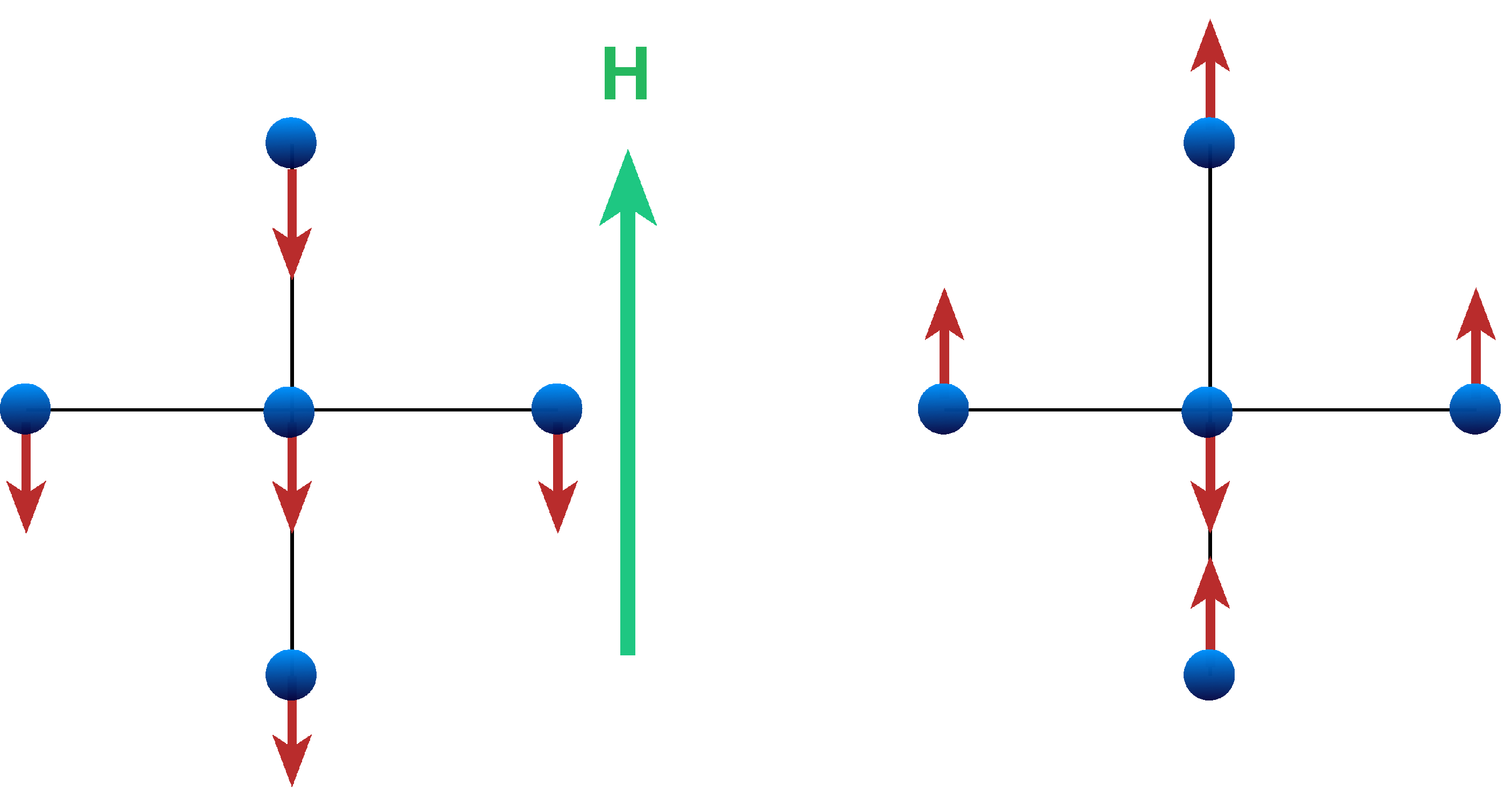}
  \centering
     \caption{
The favored situation is where a firm and its neighbors have the same status i.e. either working with maximum capacity or minimum capacity. In this figure we have considered a two dimensional lattice where each firm is connected to four other firms. In times of crisis where all firms work with minimum capacity, if the government compensates for decline of neighbors, each firm feels that neighbors work with maximum capacity. In the Ising model when neighbors are downward, if you impose an exogenous field equal to $8J$ then it is like all neighbors are upward.}\label{figising}
\end{figure}

Each manager looks at her neighbors and decides to minimize or maximize production with the probabilities as
\bea\begin{split}\label{probabilitiyising}
&P_{\uparrow}\propto \exp{\{\frac{(N_{\uparrow}-N_{\downarrow})J}{T} \}},
\cr&P_{\downarrow}\propto \exp{\{\frac{(N_{\downarrow}-N_{\uparrow})J}{T}\}},
\end{split}\eea
where $N_{\uparrow/\downarrow}$ represents the number of neighbors which work with maximum/minimum capacity and $P_{\uparrow/\downarrow}$ indicates the probability to choose a high or low level of activity.

The parameter $J$ shows the strength of connection between two firms. Actually it should be related to the purchase of neighbors from each other where for now we have supposed to be homogenous. In economy we have: the bigger the trade, the stronger the interaction. In this model, the bigger the $J$, the stronger the interaction. So, the strength of trade between firms is encapsulated in $J$. The parameter $T$ controls the stochastic behaviors. If we let $T\rightarrow 0$ then it means that managers have no stochastic behavior; i.e. if the majority of neighbors of a firm work with maximum/minimum capacity then it definitely works with maximum/minimum capacity. By comparison if we let $T$ grow comparing to $\bar N J$ where $\bar N$ is the average degree of the network, the chance for a firm to work with maximum or minimum capacity becomes equal; i.e. the behavior becomes random and correlation between neighbors tends to zero. 

If we aim to encapsulate the positive correlation between activity of nodes and the uncertainty of behaviors in a single parameter, the best candidate would be the temperature. If the interaction between firms are strong, the system would reside in the cold phase of the ferromagnetic Ising model. In such a case, if a major portion of firms work with minimum capacity, then without shocks, recovery would be unlikely. If we accept to model the real networks of firms with a homogenous Ising model still finding proper temperature is not easy. The long lasting stagnations after big downturns such as the Great Depression or the Great Recession however is a sign that in some senses the metastable states could exist in economy which for our model suggests that the temperature should be relatively low. The stagnation after the great depression lasted more than a decade leading the second world war. Once the war started and a the government purchase boosted then economy fell into the right track. Even after the war when government purchase declined economy kept its good shape. That is why in a Keynesian economics it is believed that without government stimulation, economy may live in a long lasting depression. If we believe Keynes's terminology, then we should think of temperatures below the critical temperature.

When the government aims to stimulate economy through fiscal policy, it makes some orders from the private party. Regarding networks, it tries to compensate decline of orders by nodes from each other. This means that in our dipole model most of dipoles are downward. When government makes extra purchase from the private party, for firms this new order compensates part of decline of order by their neighbors. So, in strategies in Eq. (\ref{probabilitiyising}) we should add a term similar to the external field for the role of government stimulus purchase.  

We suppose that in recession, the system lives in the vacuum where most of firms work with minimum capacity. 
Now we impose a stimulus field in an upward direction and track magnetization. We track the magnetization and measure the number of Monte Carlo steps that the stimulus field needs to change the status of at least half of the dipoles to an upward direction denoted by $\tau$. To state clearer $\tau$ is the number of Monte Carlo steps needed for stimulation to elevate the magnetization above zero. In economy a stimulus bill can be imposed in a few seasons or a couple of years. The total bill however is important. So, in our Ising model of economy we should be interested in $\tau H$.  

We can impose a relatively weak field $H$. The value for $\tau$ however increases. On the contrary we can impose a stronger field and decrease the value of $\tau$. Now, there is a question. Under what circumstance we can decrease $\tau H$? Dynamic of the Ising model and its metastability has been widely studied, see for example \cite{chakrabartih}-\cite{Rikvold} and references therein. The response of the system depending on the strength of the stimulating field are divided in stochastic an deterministic regimes. Our interest is in deterministic regime where the system changes its vacuum in a predictable time passage.

It can be shown that this value has a minimum bound where we cannot have a successful stimulation for hits smaller than such a bound. Besides, it has been shown that this value is translated to the minimum bound for successful stimulation in economy as
\bea 
bill_{min}=0.44\Delta GDP
\eea
where $\Delta GDP$ is the gap between GDP in expansion and recession. 
       
It is interesting to notify the observation that despite simplicities in the model it had successful predictions for two of the biggest economies in the world. While the US stimulus bill was above this threshold and successful, the EU bill was far below this threshold and unsuccessful, \cite{hosseinyising}.


The model is so far too simple to be reliable for the application in economy. The important point however is that it suggests that local correlation in the network provides a hysteresis and to overcome such hysteresis we need a minimum bound for successful stimulation. The major question to be answered is that what happens if we go to more realistic models. Ising model supposes agents have binary choices. 
In economy, managers can choose any level of activity between maximum and minimum capacities. 
Should we still expect a minimum bound for a successful stimulation in interactions with continuous choices such as the XY model?


\section*{The XY model}\label{secproductionfunction}

\begin{figure}[]
        \centering
  \includegraphics[width=.6\columnwidth]{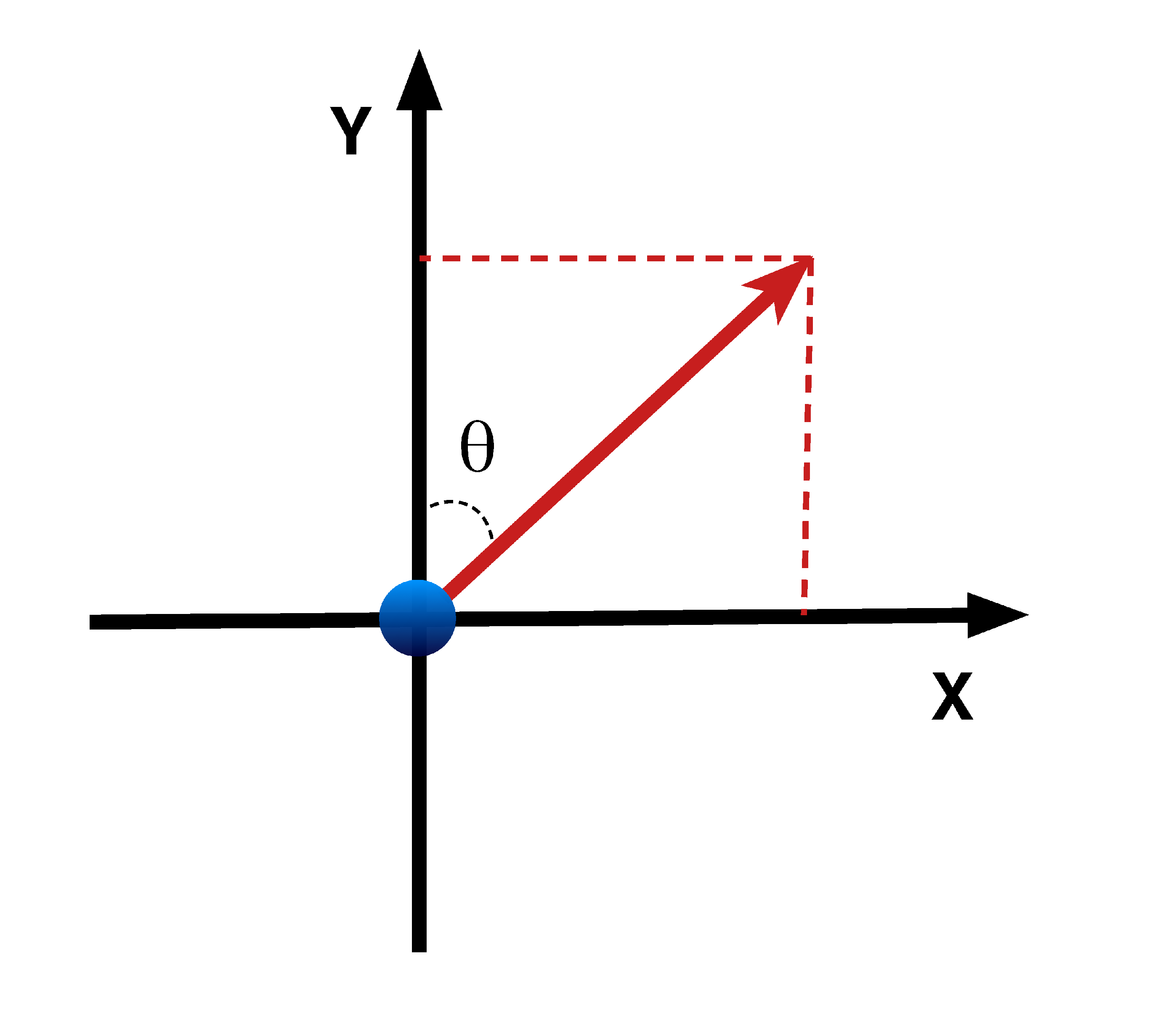}
     \caption{In the XY model of firms, the projection on $Y$ direction identifies the level of activity. If $\cos\theta_i\approx 1$ then the firm works with maximum capacity. An $\cos\theta_i\approx 0$ means a middle level of activity and $\cos\theta_i\approx -1$ identifies a low level of activity.}
     \label{xy8}
\end{figure}

To bypass restrictions of the Ising model and to be closer to the real world we consider the XY model. It is mainly because in the real world managers have much more than binary choices. In this perspective we can suppose that each manager as an agent chooses a random angle in the XY plane. We then suppose that the cosine of the angle with the positive Y direction represents the level of activity of a firm. A zero angle means the highest level of activity and an angle equal to $\pi$ means the lowest level of activity, see Figure \ref{xy8}. In other words, given the firm $i$ we can denote its level of activity by 
\bea 
Y_i=\frac{1}{2}\times(Y_{max}+Y_{min})+(Y_{max}-Y_{min})\times\cos\theta_i\;\;
\eea
where $Y_{min}$ and $Y_{max}$ represent the minimum and maximum capacity of production of the firms.

Now the XY model of firms makes sense. If neighbors of a firm work with a higher level of activity, they force the firm to work with a higher level. The intensity of their force however is a function of their angles, i.e. it is a function of their level of activities. In our Monte Carlo simulation we update directions proportional to their weights
\bea  
P(\theta_i)\propto\exp{\{-\frac{\sum_{j} Jcos(\theta_i-\theta_j)}{T}\}}
\eea     
where summation is over the neighbors of node $i$. This probability notifies one fact; though any direction might be chosen by each node, angles closer to the neighbors' one are more probable.  

In economic language this means that managers have a stochastic and heterogenous behavior. It is however more likely that they choose a level of activity closer to their neighbors. Now, we suppose that in economic crises a majority of firms work with a low level of activity. We then ask if we want to stimulate such a network to work with a higher level of activity, what would be the response of the network? 
Should there exist some resistance from the network for recovery or equivalently is there still a minimum bound for a successful stimulation?

%


\section*{Results}

We first suppose that in recession all firms work with their minimum capacity. In the XY language, all dipoles are along the negative Y direction. A fiscal stimulation aims to compensate the decline of orders. In our XY language it can be modeled by a magnetic field along the positive Y direction aiming dipoles to modify directions towards this direction. In an experiment we ran a simulation for a Watts-Strogatz network with 512 nodes, with the average degree $K=8$, and the probability for rewiring $P=0.1$. Since at the beginning all dipoles were along the negative Y direction, we had $\bar m_y=-1$. We then imposed a magnetic field with the intensity $H$ and updated the direction of dipoles under such a field. We tracked the net magnetization and measured how many Monte Carlo steps were needed for the stimulus field to change $m_y$ to elevate above zero. The value which we were interested in was $\tau H$ which in economy means the total bill imposed within a number of seasons. 

\begin{figure}[] 
        \centering
  \includegraphics[width=.8\columnwidth]{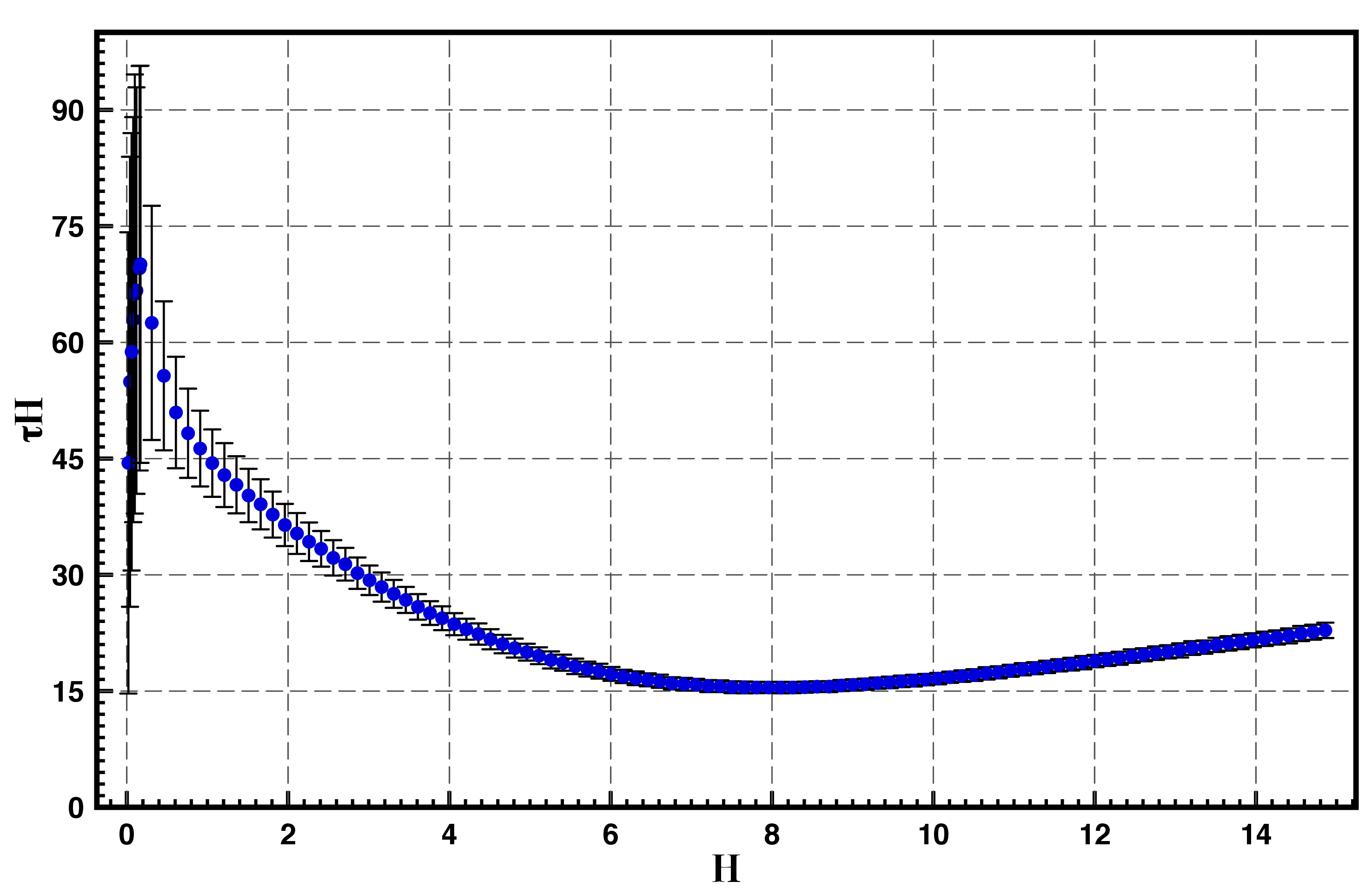}
     \caption{
 The X axis shows the intensity of the stimulus field. The Y axis shows the size of the hit or $\tau H$. As it can be seen for week fields the response of the system is stochastic where the standard variation of successful hits is comparable with its mean value. For strong intensities however the response is deterministic. The response has a minimum in the strong intensity regime.}\label{thh}
\end{figure}

The result of our simulation for an ensemble of 1000 runs is graphed in Figure \ref{thh}. As it can be seen, there are two regimes: a deterministic regime and a stochastic regime. If we impose a relatively strong field we can expect the magnetism to become zero after a predictable period. If we impose a weak field then the response of the network is stochastic, in a sense that the standard variation for $\tau H$ is comparable with its mean value. 

In the world of economy, the stochastic regime is not of interest. First of all, neither policy makers nor politicians are interested in stimulations which their outcomes are stochastic. Secondly, if you want to wait that long, some other parameters such as technological shocks may after all help economy to recover. Keynes says: "In the long run we are all dead". The aim of stimulation is not to leave economy with stochastic shocks. It is an act aiming to recover economy within a reasonable time passage. So, we focus our simulations and analysis within this region. In this region we find a minimum around $H=8$ where the minimum of $\tau H$ is $15.7\pm0.72$. We have so far a good news which is the existence of the minimum bound for successful stimulation. Existence of such a bound however is not a good news if it is not universal. In other words we should look for the minimum bound in other networks and see if it is independent of the network characteristics.  

\section*{Sensitivity Analysis }\label{sectionexamples}

\begin{figure}[]
        \centering
  \includegraphics[width=.8\columnwidth]{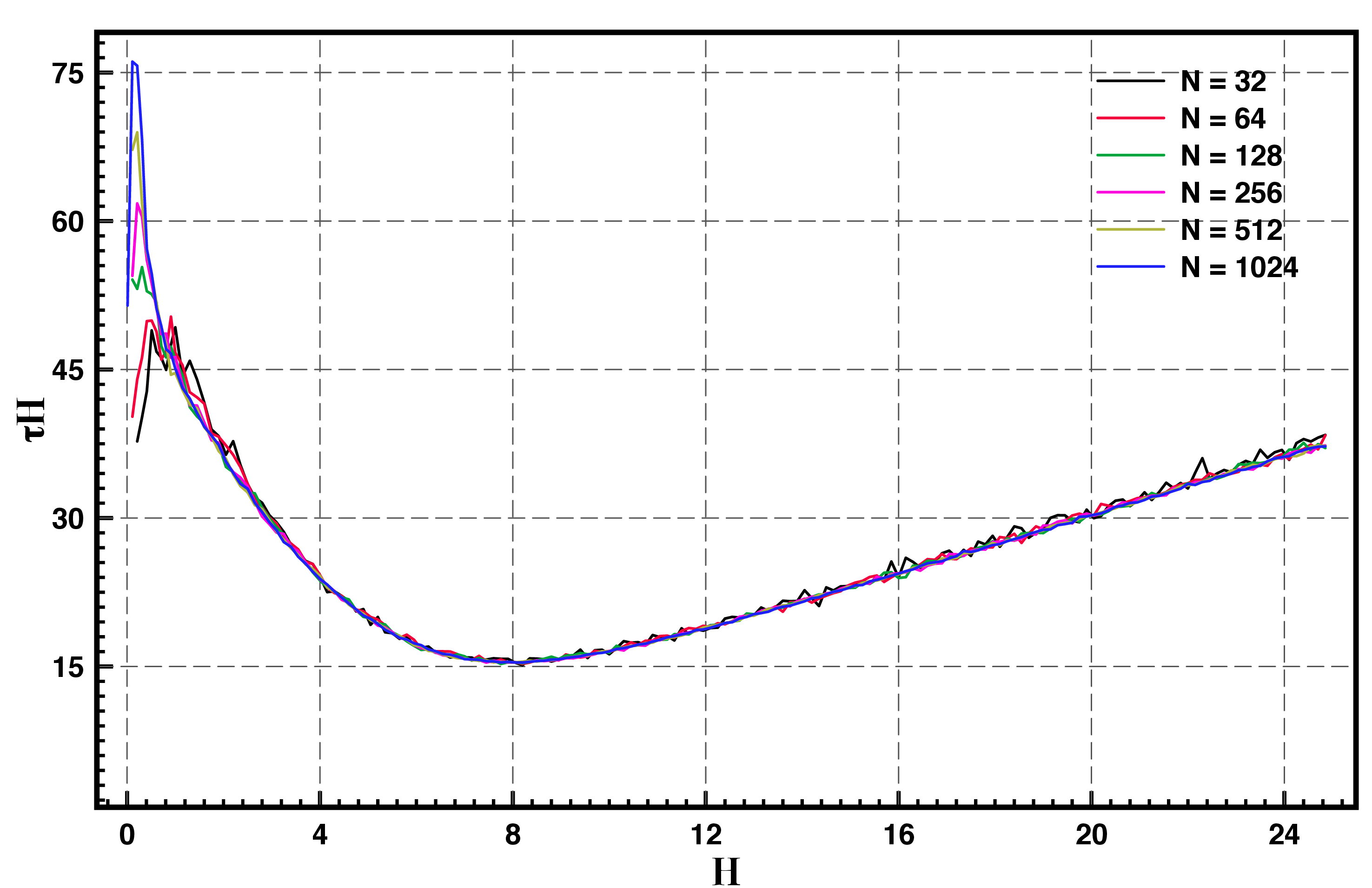}
     \caption{
 As it can be seen, in the deterministic regime, the responses of the systems are not related to their sizes.}\label{size}
\end{figure} 

We now need to check if our minimum bound depends on the properties of the network such as size or degree distribution. So we should perform some analyses. The first point to be checked is the dependency of the result on the size. In Figure \ref{size} we have depicted the result of simulation for different sizes. As it can be seen, the minimum bound of successful stimulations are independent of the sizes. This is a reasonable property. Within the regime we study, the stimulating field is strong. As a result, the metastable lifetime is short, namely a few Monte Carlo steps. So, the boundary conditions do not affect the result. 

\begin{figure}[]
\centering
  \includegraphics[width=.8\columnwidth]{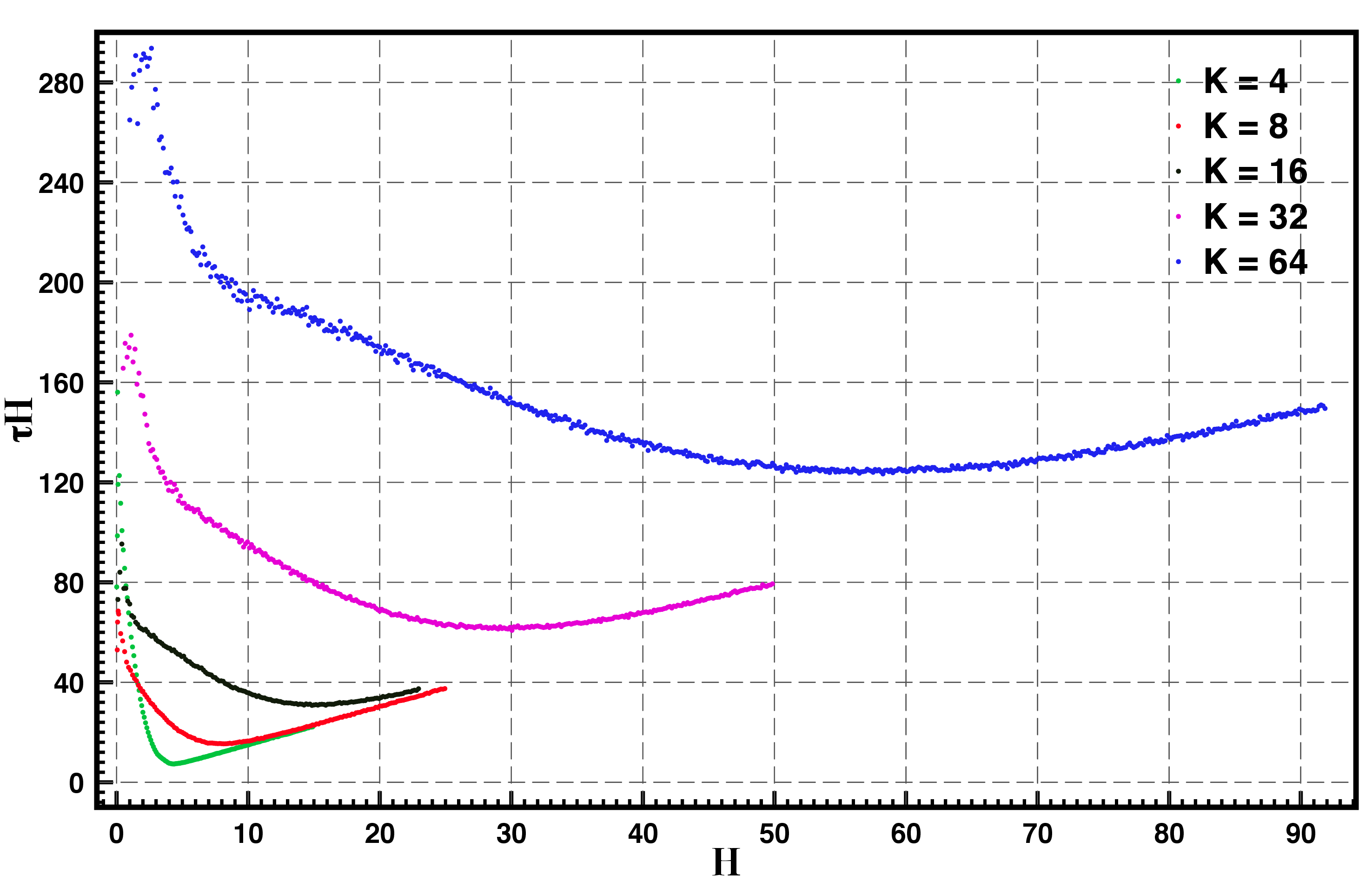}  
     \caption{
Different degree averages suggest different curves for the response of the system to the external field}\label{k}
\end{figure}

\begin{figure}[]
        \centering 
  \includegraphics[width=.8\columnwidth]{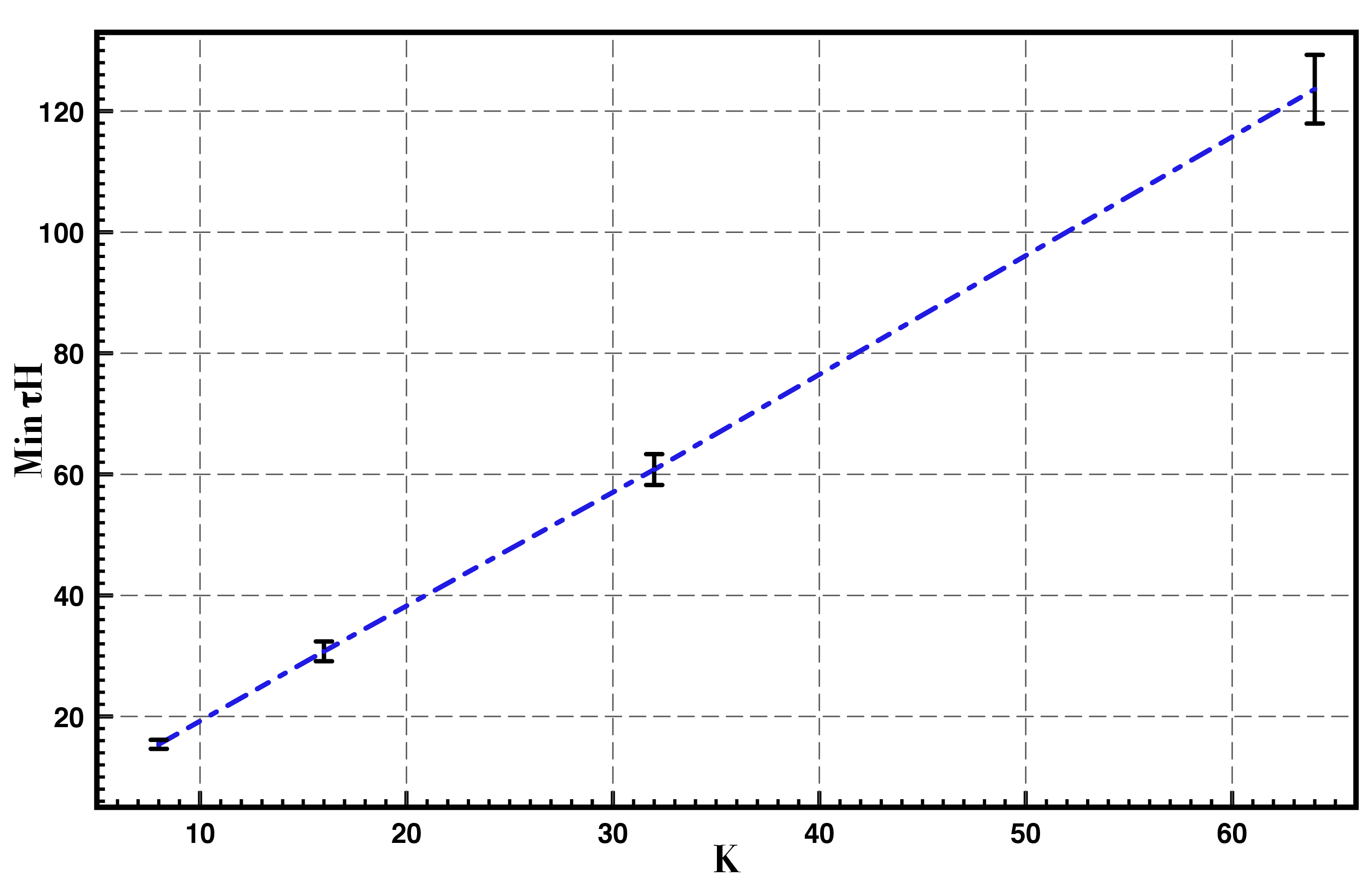}
     \caption{
 Minimum of $\tau H$ for different values of average degrees in Figure \ref{k} has been graphed. As it can be seen, the minimum of $\tau H$ grows linearly with K.}\label{knormal}
\end{figure}

The second point to be checked is the dependency of the result to the average degree. The results are presented in Figure \ref{k}. From the graph it is clear that the minimum bound grows as the average degree grows. Now we graph the minimum bound vs the average degree in Figure \ref{knormal}. As it can be seen, the minimum grows linearly with the  average degree. this means that our minimum bound is related to a property of the networks. If we make a wise analysis however this property is an advantage. 

Consider a crisis where a major portion of firms work with minimum capacity. In this case in our dipole models a major of nodes align towards the negative Y direction. Now if the stimulus bill is big enough to compensate the decline of orders, the firms would not mind about any decline of orders by neighbors. In this case, in average for each firms it looks like that its neighbors are working with maximum capacity. In the XY model and in a Watts-Strogatz network with average degree $K$, we suppose that all nodes are downward. If a stimulus field has an intensity equal to $2K$ in the positive Y direction, then for each dipole it is equivalent to the cases where neighbors are upward. So, the gap between production in expansion and depression is proportional to the average degree. Then we can write
\bea\label{timestep} 
\Delta GDP\propto 2KJ
\eea  
in which J is interaction in the XY model, and $\Delta GDP$ is the gap between expansion and depression. To have a quantitative result we need to identify other important parameters. 

Actually the missing object in our discussion is the link between time in our model and the real world. What is a Monte Carlo time step in the real world? The fact is that for different sectors of economy we should have different time steps. For those sectors which need unskilled or low skilled employees it is much easier to reduce or increase production. For sectors with professional employees it would be harder to fire \& hire employees, since the   instability caused by this action is more expensive. For some other sectors such as construction, when a project is started it would be hard to abandon. So, the overall conclusion is that the time step for a manager to change strategy for the level of activity is heterogenous. For the simplest case however we can suppose to have similar time steps. The majority of contracts between employees and employers are on annual basis. So, if a manager aims to reduce production to its minimum capacity, in average she should wait 6 months for contracts to be fulfilled. This means that for the simplest case we can suppose that one Monte Carlo step is six months.

If we suppose a Monte Carlo step in an XY model is six months, then we can rewrite equation (\ref{timestep}) as
\bea 
\Delta GDP\approx4KJ.
\eea
This equation states that if in an XY model we impose a magnetic field with a strength twice the interaction between nodes before updating all the nodes, then it is equivalent with the case where we have stimulated economy with the gap for aggregate production in six months. Now if we stimulate our XY model with different strengths and different numbers of Monte Carlo steps, then for the related bill in the economy side we can write 

\bea\label{billrelation} 
\frac{bill}{\Delta GDP}\approx\frac{\tau H}{4KJ}.
\eea
 This equations is very flash. It states that if the value of $\tau H$ in the XY side grows linearly with K, then in the economy side the related stimulation is independent of K. So, our minimum bound is independent of the average degree.


\begin{figure}[]
        \centering 
  \includegraphics[width=.8\columnwidth]{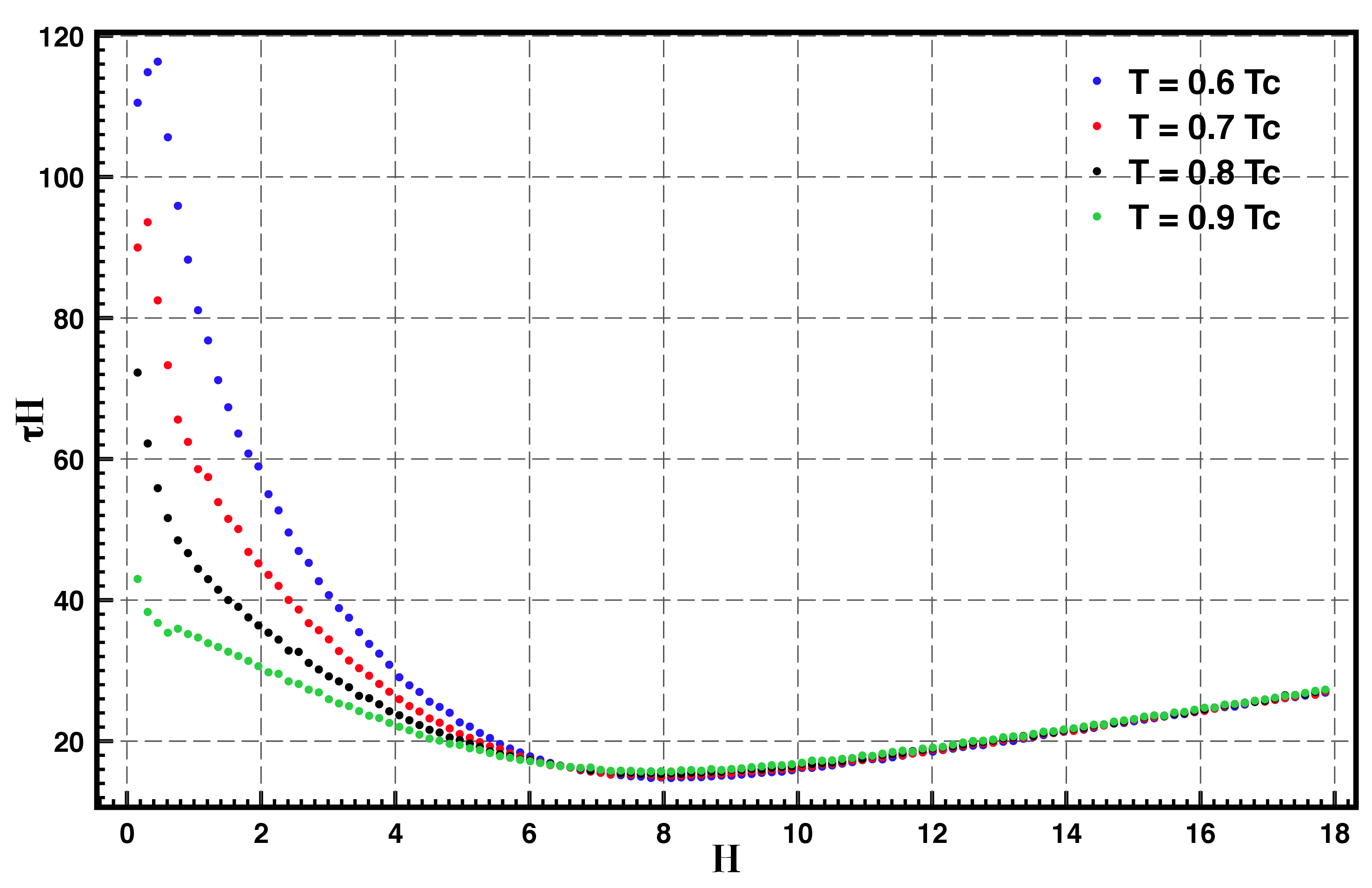}
     \caption{
As it can be seen, the minimum of $\tau H$ is not seriously influenced by temperature.}\label{temperature}
\end{figure}

Another point to be checked is the relation between the minimum bound and temperature. In another simulation for a network with 500 nodes and $K=8$ we performed simulation for a range of temperatures below $T_c$. The result has been depicted in Figure \ref{temperature}. As it can be seen, the minimum of $\tau H$ is not substantially affected by the temperature. So, our result in a reasonable range of temperature would not depend on the temperature itself. 

\begin{figure}[]
        \centering 
  \includegraphics[width=.8\columnwidth]{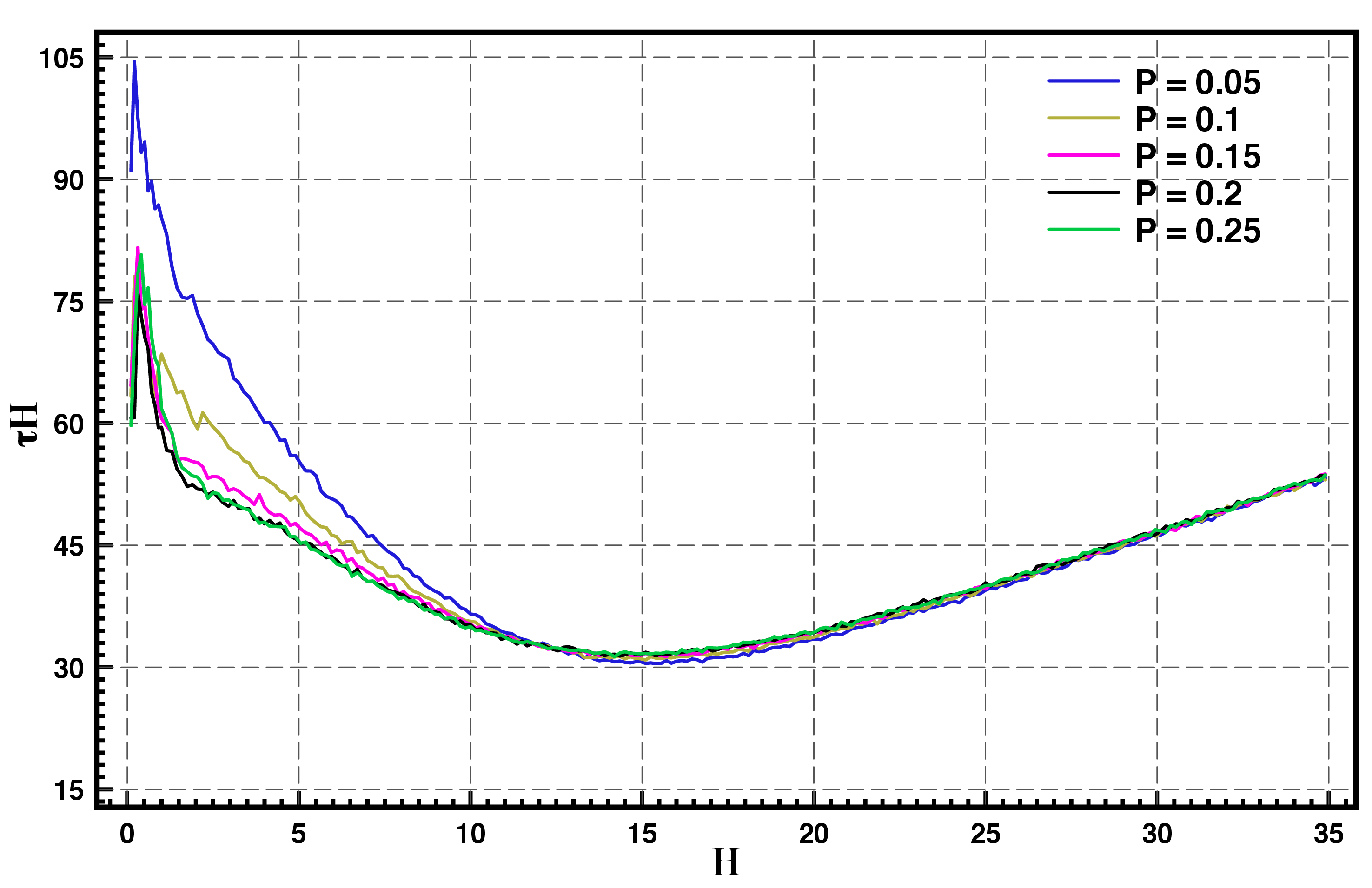}
     \caption{
For a Watts-Strogatz network the degree average has been kept equal to 16. The probability for reconfiguration (P) in the network has been changed within a range of 0.05 to 0.25. As it can be seen, the minimum of $\tau H$ is independent of $P$.}\label{p}
\end{figure}

As another analysis we checked the impact of the value of rewiring probability $P$. We changed its value in the network and performed simulation for $T=0.8 T_c$. The result has been depicted in Figure \ref{p}. As it can be seen, again the minimum bound is independent of the network characteristics.

So far we have supposed that recession can be represented by a network with all nodes downward. One may however argue that in recession not all firms work with their minimum capacity. In this case we need to change our initial conditions. So we perform another experiment.

In this experiment, we supposed that in bubbles and expansions before the crises, a big portion of firms work with their maximum capacity. Deep crises is denoted by the situation where a big fraction of firms reduce their production. To simulate such a situation, we first set all dipoles along the positive Y direction. We then imposed a strong downward magnetic field. Under this field some dipoles take random directions where downward directions become preferred. We tracked the magnetism along Y direction until it became $m_y=-0.5$. This means that some firms were working with their higher capacities. Some firms were working with lower capacities and some were in the middle. The overall result however represents a serious decline in production. We suppose that it resembles a situation close to the situation in downturn such as the crash of economy after the Lehman Brothers bankruptcy.

\begin{figure}[]
        \centering
  \includegraphics[width=.65\columnwidth]{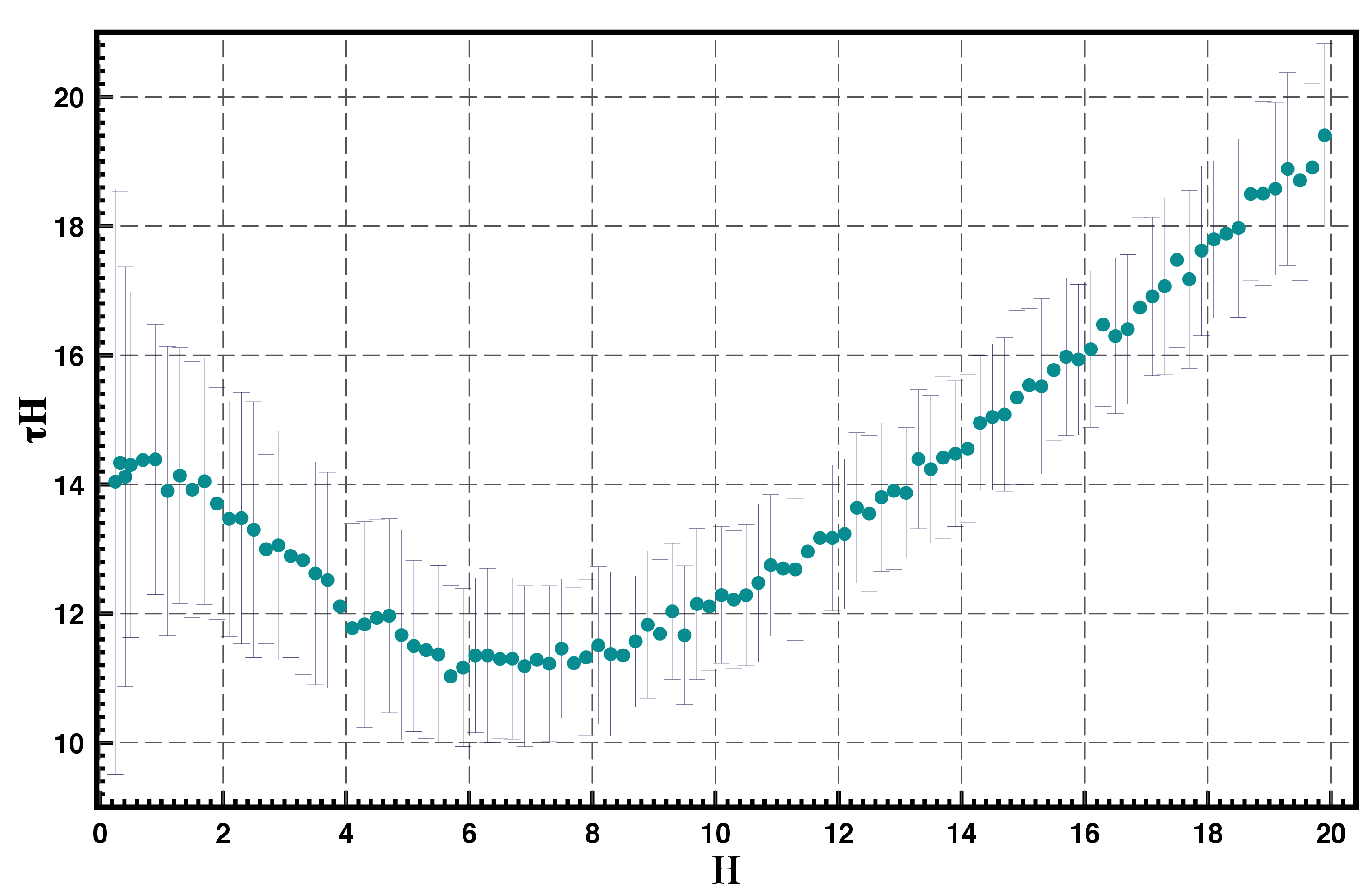}
     \caption{The response of the system were in the initial condition magnetization is $m_y\approx -0.5$. For this initial condition the minimum value of $\tau H$ is equal to $11.2\pm1.16$}\label{random}
\end{figure}

We now, impose an upward magnetic field and check its influence. The result has been depicted in Figure \ref{random}. As it can be seen, still the minimum bound for stimulation exists. This bound however has been dropped from our pervious situations to $\tau H_{min}=11.2\pm1.16$. It however should be noticed that if we suppose that in a recession magnetization is around $m_y=-0.5$ then we should rewrite equation (\ref{billrelation}) as
\bea 
\frac{bill}{\Delta GDP}\approx\frac{\tau H}{3KNJ}.
\eea 
This is because, in this case the gap for GDP belongs to the situations where we have $m_y=-0.5$ and $m_y=1$. This equation suggests the minimum bound for stimulation to be $\tau H_{min}=(0.47\pm0.05)\Delta GDP$. Surprisingly we observe that still the minimum bound for successful stimulation is not changed substantially. 

\section*{Conclusions}

Physicists have tackled a wide spectrum of interdisciplinary problems which ranges from 
biology \cite{Peng}-\cite{Mantegna94} to social sciences \cite{Nekovee}-\cite{Yasseri}, econophysics \cite{farmer}-\cite{Aoyama2010} and many other areas of researches. 
One of the major concerns in such areas of researches is the matter of time. While in statistical physics, regularly time is not of interests and phenomena are studied at equilibrium, for a big portion of systems such as those studied in socioeconomics, time is a crucial parameter.
This difference results in some unexpected effects.

Ergodic theory for example claims that if a finite size system is left isolated, it spans almost the entire phase space and bypasses different equilibria. Ergodic theory works for a great range of phenomena in physics where we usually have enough time for evolution. In nature and in societies however time is a matter of survive. So, we may not be able to wait for a long time until various local equilibria of the phase space are met. In such cases forcing the system to move to our desired part of the phase space is not costless.


Though theoretically both Ising and XY models have no hysteresis, when we are concerned with time we observe metastability and resistance. Metastability and dynamic hysteresis of the Ising model has been widely studied in the literature. In this paper however we focused on the XY model. 


Despite the Ising model, in the XY model below the critical temperature we have a continuous range of vacuums. Though these vacuums share the same level of energy, still triggering the system to the desired vacuum has a minimum cost which is an interesting observation. When translated into our agent based model of economy we observe that in a model more realistic than the Ising model, still there is a dynamic hysteresis and to overcome it we need a minimum bound of budget. This minimum is about $0.48\Delta GDP$ where $\Delta GDP$ is the gap for GDP between recession and expansion. 

It should be notified that as the major goal of the work we showed that if agents have the choice to put production level to a continous level, still the minimum bound for stimulation exists. This means that the minimum bound suggested for successful stimulation is not the result of simplification of the economic networks to the Ising model. Beside this finding we showed that the minium bound is close to our earlier observation. Despite our prior goal this finding is not a concrete result. Actually, to translate our result to the economic values, we supposed that one Monte Carlo step means one year in the Ising model and six months in the XY model. One should notify that in real world, managers have choices such as declining the working hours. Such decision can be made pretty fast. To have a proper interpretation for the economic world we need to bring such possibilities into account. In such cases though the minimum bound exist, its value will be less than the prediction of the Ising model.

In our model we supposed that interactions for all nodes are positive. This is while in the real world some interactions are negative. If Ford sells more, then there is a good chance that Chevrolet sells less. So, an improvement of the analysis could be an extension to the spin glass like model. Moreover, for more realistic cases a heterogenous network with different weights should be considered. 
 
Emergence is the most important issue discussed in complex systems. It arises when we study the collective behavior of many body systems. One of the major goals of econophysicists is to understand the collective behaviors in economy, see for example \cite{hosseinypercolation}-\cite{hosseinyrole}.

Though we were concerned with the economic networks, metastability and dynamic hysteresis may arise in many other networks and social systems. An example is the voting model. If people can influence each other's attitudes, then social paradigms may live in local equilibria.  
Studying metastability of such systems will be harder than economical systems, but more exotic.  
Actually in any game theoretical model, dynamic hysteresis may exist when we aim to move from a local equilibrium to another one. 




\begin{thebibliography}{10}
\bibitem{Schweitzer}
Frank Schweitzer, Giorgio Fagiolo, Didier Sornette, Fernando Vega-Redondo, Alessandro Vespignani, Douglas R White, "Economic networks: The new challenges", Science 325, 5939, 422-435, (2009) 


\bibitem{battiston}
S Battiston, DD Gatti, M Gallegati, B Greenwald, JE Stiglitz,"Liaisons dangereuses: Increasing connectivity, risk sharing, and systemic risk", Journal of Economic Dynamics and Control 36 (8), 1121-1141


\bibitem{lenz}
Lenzu, S. \& Tedeschi, G.
Systemic risk on different interbank network topologies. Physica A: Statistical Mechanics and its Applications \textbf{391(18)}, 4331-4341 (2012).

\bibitem{contreras}
Contreras, M. G. A. and  Fagiolo, G. "Propagation of Economic Shocks in Input-Output Networks: A Cross-Country Analysis." Phys. Rev. E. \textbf{90}, 062812 (2014).

\bibitem{Acemoglu}
 Acemoglu, D., Carvalho, V., Ozdaglar, A. $\&$ Tahbaz-
Salehi, A. The network origins of aggregate fluctuations. Econometrica \textbf{80(5)}, 1977-2016 (2012).

\bibitem{krugmanend}
Krugman, P. End This Depression Now! (WW Norton $\&$ Company, NY ,2012)


\bibitem{stiglitzfreefall}
Stiglitz, J. Freefall$\:$ America, free markets, and the sinking of the world economy (WW Norton $\&$ Company, NY ,2012)




\bibitem{hosseinyising}
Hosseiny, A., Bahrami, M., Palestrini, A. \&  Gallegati, M. Metastable Features of Economic Networks and Responses to Exogenous Shocks. PLoS ONE. \textbf{11(10)}, e0160363 (2016).


\bibitem{William}
  Brock, W. A. $\&$ Durlauf, S. N. Discrete choice with social interactions. The Review of Economic Studies \textbf{68(2)}, 235-260 (2001).


\bibitem{Durlauf}
  Durlauf, S. N. $\&$ Ioannides, Y. M. Social interactions. Annu. Rev. Econ. \textbf{2(1)}, 451-478 (2010).










\bibitem{Zhou}
WX Zhou, D Sornette, "Self-organizing Ising model of financial markets"
The European Physical Journal B 55 (2), 175-181

\bibitem{chakrabartih}
Bikas K. Chakrabarti and Muktish Acharyya, "Dynamic transitions and hysteresis", Rev. Mod. Phys. 71, 847 (1999)



\bibitem{Rikvold}
Rikvold, P.A. Tomita, H., Miyashita, S., Sides, S.W. "Metastable lifetimes in a kinetic Ising model: dependence on field and system size." Physical Review E 49.6 (1994): 5080.

\bibitem{Peng}
Peng, C.K., Havlin, S., Stanley, H. E., Goldberger, A. L. Quantification of scaling exponents and crossover phenomena in nonstationary heartbeat time series. Chaos: An Interdisciplinary Journal of Nonlinear Science. \textbf{5(1)}, 82-87 (1995).



\bibitem{Friedrich}
Friedrich, R., Peinke, J., Sahimi, M. \& Tabar, M.R.R. Approaching complexity by stochastic methods: From biological systems to turbulence. Physics Reports. \textbf{506(5)}, 87-162 (2011).



\bibitem{Mantegna94}
Mantegna, R. N. \emph{et~al.} Linguistic features of noncoding DNA sequences.
Physical review letters. \textbf{73(23)}, 3169 (1994).



\bibitem{Nekovee}
Nekovee, M., Moreno, Y., Bianconi, G. \& Marsili, M. Theory of rumour spreading in complex social networks. Physica A: Statistical Mechanics and its Applications, \textbf{374}, 457-470 (2007).



\bibitem{Yasseri}
Yasseri, T., Sumi, R., Rung, A., Kornai, A. \& Kert\'esz, J. Dynamics of conflicts in Wikipedia. PLoS ONE \textbf{7(6)}, e38869 (2012).




\bibitem{farmer}
  Farmer, J. D. \emph{et~al.} A complex systems approach to constructing better models for managing financial markets and the economy. The European Physical Journal Special Topics \textbf{214(1)}, 295-324 (2012).

   
\bibitem{preis}
 Preis, T., Kenett, D. Y., Stanley, H. E., Helbing, D., \& Ben-Jacob, E. Quantifying the behaviour of stock correlations under market stress. Sci. Rep. \textbf{2}, 752 (2012). 

\bibitem{preis}
Cerqueti, R, Ausloos, M. Evidence of economic regularities and disparities of Italian regions from aggregated tax income size data, Physica A: Statistical Mechanics and its Applications 421, 187-207, (2015) 

\bibitem{mastromatteo}
 Mastromatteo, I., Toth, B. \& Bouchaud, J. P. Agent-based models for latent liquidity and concave price impact. Phys. Rev. E \textbf{89}, 042805 (2014).


\bibitem{Safdari}
Safdari, H., Hosseiny, A., Farahani, S. V. \& G. R. Jafari A picture for the coupling of unemployment and inflation. Physica A: Statistical Mechanics and its Applications \textbf{444}, 744-750 (2016).

\bibitem{Scalas}
Scalas, E., Gorenflo, R. \&  Mainardi, F. Fractional calculus and continuous-time finance. 
Physica A: Statistical Mechanics and its Applications \textbf{284 (1)}, 376-384 (2000).



\bibitem{hosseinygeommetry}
Hosseiny, A. Geometrical Imaging of the Real Gap Between Economies of China and the United States. Physica A: Statistical Mechanics and its Applications \textbf{479}, 151-161 (2017).



\bibitem{Dragulescu}
Dragulescu, A. \& Yakovenko, V. M. Statistical mechanics of money. The European Physical Journal B-Condensed Matter and Complex Systems \textbf{17}, 723-729 (2000).



\bibitem{Aoyama2010}
Aoyama, H., Fujiwara, Y., Ikeda, Y., Iyetomi, H. \& W. Souma Econophysics and Companies: Statistical Life and Death in Complex Business Networks 234 (Cambridge University Press, Cambridge, 2010). 


\bibitem{hosseinypercolation}
Shirazi, A., Saberi, A.A., Hosseiny, A., Amirzadeh, E., Simin, P.T., "Non-criticality of interaction network over system's crises: A percolation analysis", Scientific reports 7 (1), 15855, (2017)

\bibitem{hosseinyrole}
Hosseiny, A., Gallegati, M., "Role of intensive and extensive variables in a soup of firms in economy to address long run prices and aggregate data"
Physica A: Statistical Mechanics and its Applications 470, 51-59, (2017)



\end{thebibliography}
\end{document}